\documentclass[final,5p,times,twocolumn,sort&compress]{elsarticle}

\usepackage{amssymb}

\usepackage{graphicx}
\usepackage{dcolumn}
\usepackage{bm}
\usepackage{cprotect}

\usepackage[normalem]{ulem} 
\newcommand\redout{\bgroup\markoverwith
{\textcolor{red}{\rule[.5ex]{2pt}{1pt}}}\ULon}
\usepackage{xcolor}

\journal{Physica A}

\begin{document}

\begin{frontmatter}



\cprotect\title{Global exploration of phase behavior in frustrated Ising models using unsupervised learning techniques}


\author[inst1]{Danilo Rodrigures de Assis Elias}

\affiliation[inst1]{organization={Instituto de F{\'i}sica Gleb Wataghin, Universidade Estadual de Campinas, UNICAMP},
            city={Campinas},
            postcode={13083-859}, 
            state={S{\~a}o Paulo},
            country={Brazil}}
            
\author[inst2]{Enzo Granato}
\author[inst1,inst3]{Maurice de Koning}

\affiliation[inst2]{organization={Laborat{\'o}rio Associado de Sensores e Materiais, Instituto Nacional de Pesquisas Espaciais},
            city={S{\~a}o Jos{\'e} dos Campos},
            postcode={12227-010}, 
            state={S{\~a}o Paulo},
            country={Brazil}}

\affiliation[inst3]{organization={Center for Computing in Engineering \& Sciences, Universidade Estadual de Campinas, UNICAMP},
            city={Campinas},
            postcode={13083-861}, 
            state={S{\~a}o Paulo},
            country={Brazil}}

\begin{abstract}
We apply a set of machine-learning (ML) techniques for the global exploration of the phase diagrams of two frustrated 2D Ising models with competing interactions. Based on raw Monte Carlo spin configurations generated for random system parameters, we apply principal-component analysis (PCA) and auto-encoders to achieve dimensionality reduction, followed by clustering using the DBSCAN method and a support-vector machine classifier to construct the transition lines between the distinct phases in both models.  The results are in very good agreement with available exact solutions, with the auto-encoders leading to quantitatively superior estimates, even for a data set containing only 1400 spin configurations. In addition, the results suggest the existence of a relationship between the structure of the optimized auto-encoder latent space and physical characteristics of both systems. This indicates that the employed approach can be useful in perceiving fundamental properties of physical systems in situations where \emph{a priori} theoretical insight is unavailable. 
\end{abstract}


\begin{highlights}
\item We employ a machine learning approach based on unsupervised learning techniques to obtain a global picture of phase diagram from single data set. 
\item Resulting critical lines are in very good agreement with available exact results.
\item Results indicate relation between structure of dimensionality-reduced latent space and fundamental physical system properties. 
\end{highlights}


\end{frontmatter}


\section{Introduction}
\label{Introduction}

Over the past few years machine learning (ML) has revolutionized the way in which the behavior of complex systems is investigated, providing a data-driven approach that exploits the pattern-recognition powers of such techniques~\cite{Bishop2006}. In particular, it has had a tremendous impact on the physical sciences, where ML methods have been applied to a wide variety of problems originating from areas as diverse as condensed-matter and statistical physics, particle physics, cosmology, quantum computing, chemistry and materials science~\cite{Carleo2019}.

Within the field of condensed-matter physics, a major purpose of the application of ML techniques has been to discover the phase behavior of different physical systems~\cite{Carrasquilla2017,Hu2017,Casert2019,Acevedo2021,Wang2016,Chng2017,Ponte2017,Nieuwenburg2017,Deng2017,Wetzel2017,Wetzel2017a,Chng2018,Liu2018,Nieuwenburg2018,Kim2018,Mills2018,Venderley2018,Ceriotti2019,Rodriguez-Nieva2019,Freitas2020}. In this context, a variety of classical spin systems ~\cite{Carrasquilla2017,Wang2016,Kim2018,Mills2018,Hu2017,Wetzel2017,Wang2017,Wang2018a,Theveniaut2019,Acevedo2021,Alexandrou2020} have played a particularly prominent role, displaying the promise of ML techniques in the discovery of complex phases of matter from raw sampling data.

In practice, many of these applications have focused on examining phase behavior as a function of a single parameter, often temperature, seeking to characterize transitions for given fixed values of possible other model parameters. Such an approach has shown to be successful, for instance, to quantify particular critical-temperature values in a number of such spin systems~\cite{Carrasquilla2017,Hu2017,Casert2019,Alexandrou2020}.  

A different kind of application of ML techniques is to employ their pattern-recognition capabilities to gain insight into a system's \emph{global} phase behavior. In this case, instead of focusing on a single critical value for a restricted set of parameter values, the purpose now is to analyze a data set consisting of configurations generated for the full system parameter space, with the goal of estimating entire critical lines and constructing a picture of the phase diagram as a whole. Such a capability is useful, for instance, when no \emph{a priori} insight into the location of critical regions is available. Casert and co-workers~\cite{Casert2019} employed such a scheme for the active Ising model, constructing gas-liquid coexistence lines in the density-temperature parameter space. Their approach is based on a two-step procedure in which unsupervised learning techniques are first used to determine phase boundaries for a fixed value of the density, followed by a supervised learning step in which phase boundaries can be predicted for different densities. Acevedo \emph{et al.}~\cite{Acevedo2021} reconstructed the critical line between the antiferromagnetic and paramagnetic phases of the two-dimensional frustrated antiferromagnetic Ising model on the honeycomb lattice using anomaly detection~\cite{Purnomo2019} based on convolutional autoencoders. For this purpose, they trained an autoencoder on temperature-dependent data for a particular fixed value of the frustration coupling constant and then used it to analyze data generated for other values to detect the transition between ordered and paramagnetic phases using anomaly detection.

In the present paper we consider a different global learning approach in which a single unsupervised learning procedure based on dimensionality reduction is applied to a data set containing samples generated for the entire parameter space. Moreover, these configurations correspond to random parameter values so as to obtain an unbiased data set, presuming no prior knowledge of the system. As an illustration, we construct the phase diagrams of two classical lattice spin systems involving several parameters, namely, the piled-up dominoes (PUD) and zig-zag dominoes (ZZD) models~\cite{Andre1979}. These models were introduced in the 1970's as generalizations of the totally frustrated 2D Ising model~\cite{Villain1977} and incorporate effects of geometrical frustration by the existence of two different spin-spin coupling-parameter values distributed according to regular patterns on the 2D square lattice. The phase behavior of these models is nontrivial, depending on both temperature as well as the relative values of the two coupling constants. In particular, the PUD model  displays three types of second-order phase transition: two of them occurring at finite temperature between a paramagnetic phase and either a ferromagnetic or striped antiferromagnetic phase, and a third featuring a transition with vanishing critical temperature. 

We analyze the raw Monte Carlo (MC) spin-configuration data generated for randomly selected points in the models' parameter spaces by means of a three-stage data approach that consists of, (i) dimensionality reduction using principal-component analysis (PCA) and auto-encoders, (ii) clustering using the density-based spatial clustering of applications with noise (DBSCAN) algorithm, and (iii) classification using a support-vector machine (SVM). The obtained results enable us to construct the critical lines of the PUD and ZZD models and, given the availability of exact results for both systems~\cite{Andre1979}, the quality of these estimates can be assessed. 

The remainder of the manuscript has been organized as follows. In Sec.~\ref{Models} we define the PUD and ZZD spin-model Hamiltonians and describe the geometric distribution of the two coupling-parameter values across the 2D square lattice. Subsequently, we discuss the details of the employed methodology in Sec.~\ref{Methodology}, describing the MC procedure employed to generate the set of spin configurations used in the analysis, as well as the ingredients of the ML approach used to process the data. The results are presented and discussed in Sec.~\ref{Results} and we end with concluding remarks in Sec.~\ref{Conclusions}.

\section{Models}
\label{Models}
 As for the standard 2D square Ising system, the PUD and ZZD models are defined by classical spins $s_i=\pm1$ arranged on a square lattice with nearest-neighbor interactions. However, unlike the standard Ising model, the PUD and ZZD models are characterized by varying interaction-strength parameters. Specifically, the total energy of both models can be written as
 \begin{figure}[t!]
    \centering
    \includegraphics[width=1\columnwidth]{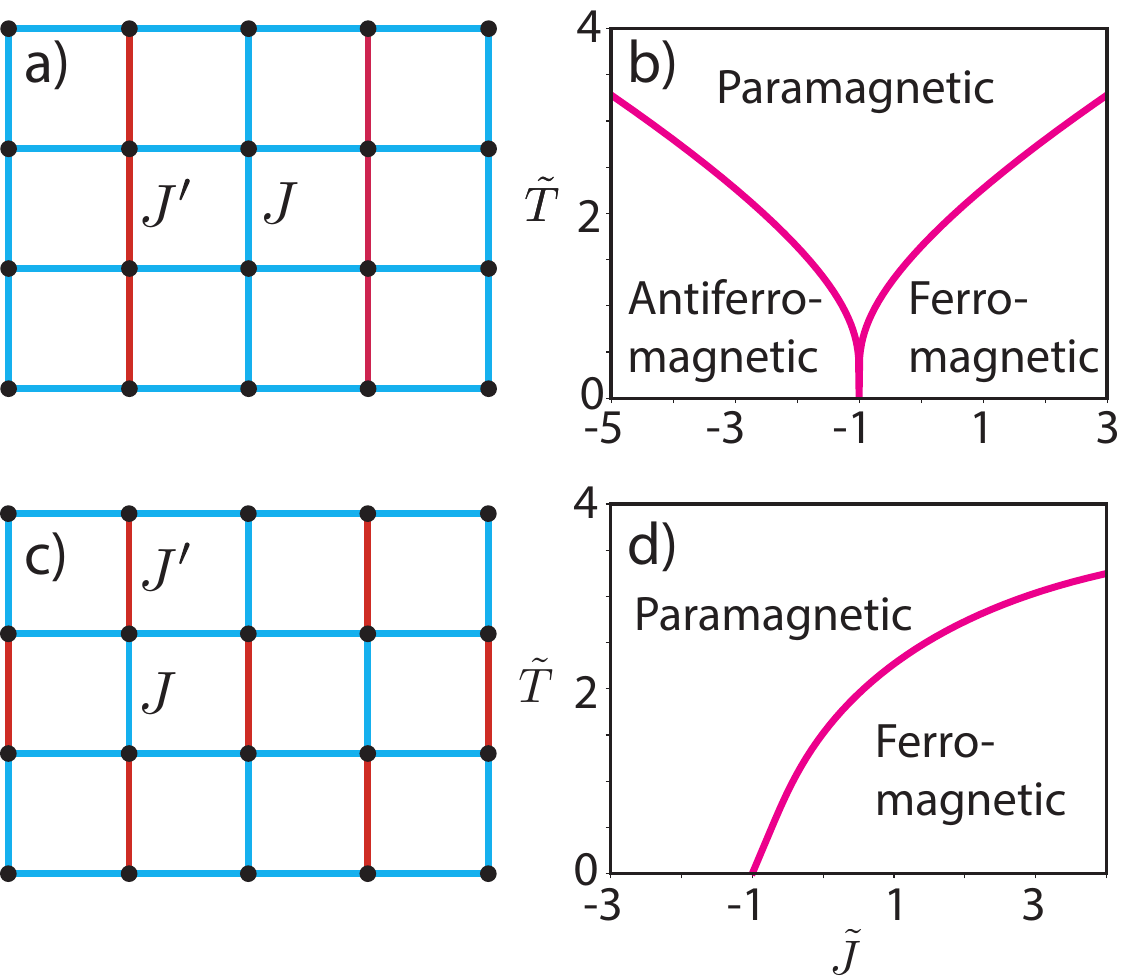}
    \cprotect\caption{Definition of spin-interaction patterns in, a), the piled-up dominoes (PUD) model and, c), zig-zag dominoes (ZZD) model as described in Ref.~\citenum{Andre1979}. Black circles represent spin sites. Blue and red links represent interactions with strengths $J$ and $J^{\prime}$, respectively. Corresponding exact phase diagrams characterized by critical lines given by Eqs.~(\ref{eq2}) and ~(\ref{eq3}) are depicted in b) and d), respectively.}
    \label{Fig1}
\end{figure}
\begin{equation}
\label{TotalEnergy}
H = -\sum_{\langle i j \rangle} J_{ij} s_i s_j,
\end{equation}
where $i$ and $j$ label the spin sites, the notation $\langle i j \rangle$ implies a summation over nearest-neighbor spin pairs and $J_{ij}$ is a spin-pair-dependent interaction-strength parameter with the dimensions of energy. For the PUD and ZZD models $J_{ij}$ can assume only two values, $J$ and $J^{\prime}$, which can be either positive or negative. In addition, their distribution across the 2D square lattice is specified in a geometrically ordered pattern, as illustrated in Fig.~\ref{Fig1}. All horizontal spin pairs interact through the coupling parameter $J$. In contrast, the interaction parameter $J^{\prime}$ couples only spin pairs that are \emph{vertical} neighbors, but only a limited set of them, with the remainder being coupled by $J$. Specifically, for the PUD model, $J^{\prime}$ acts on all pairs of alternating vertical rows, as shown in Fig.~\ref{Fig1}~a), meaning that all spin pairs in even (odd) vertical rows interact through $J
^{\prime}$, whereas all pairs in the odd (even) vertical rows are coupled through $J$. In the ZZD model, on the other hand, the interaction between spin pairs from the vertical rows alternates between $J$ and $J^{\prime}$ and such that neighboring vertical rows are ``out of phase'', creating the zig-zag pattern depicted in Fig.~\ref{Fig1}~c).   
Given the structure of the Hamiltonian in Eq.~(\ref{TotalEnergy}) the phase behavior of both models can be characterized entirely in terms of the two adimensional parameters $\tilde{J}\equiv J^{\prime}/J$ and $\tilde{T} \equiv k_B T/J$~\cite{Andre1979}, with $T$ the absolute temperature and $k_B$ Boltzmann's constant. Of course, both models reduce to the standard 2D square Ising model for $\tilde{J}=1$. Furthermore, for $\tilde{J}=-1$, the models correspond to the fully frustrated Ising model~\cite{Villain1977}. Their phase behavior is known exactly~\cite{Andre1979}, as displayed in Figs.~\ref{Fig1}~b) and d). Specifically, the PUD model features three distinct phases, paramagnetic, anti-ferromagnetic and ferromagnetic, separated by continuous phase transitions described by two critical lines that are solutions of the equations~\cite{Andre1979}
\begin{equation}
\label{eq2}
\sinh\left(\frac{2}{\tilde{T}}\right)\,\sinh\left(\frac{1+\tilde{J}}{\tilde{T}}\right)=\pm 1,
\end{equation} 
respectively, with the minus sign corresponding to the left branch. The ZZD model, on the other hand, is characterized by paramagnetic and ferromagnetic phases, separated by the critical manifold given by the solution of the equation~\cite{Andre1979}
\begin{equation}
\label{eq3}
2\tanh\left(\frac{2}{\tilde{T}}\right)\, \tanh\left(\frac{1+\tilde{J}}{\tilde{T}}\right)=1.
\end{equation}
Interestingly, the disordered paramagnetic phase persists up to zero temperature for values of $\tilde{J}$ below -1.

\section{Methodology}
\label{Methodology}

\subsection{Data Generation}
\label{DataGeneration}

The parameter space of the models is sampled randomly, employing uniform distributions for the parameters $\tilde{J}$ and $\tilde{T}$ within the intervals $\tilde{J} \in (-3,1)$ and $\tilde{T} \in (0,3)$ for the PUD model and $\tilde{J} \in (-3,3)$ and $\tilde{T} \in (0,4)$ for the ZZD system. To this end we fix $J=1$ and sample $J^{\prime}$ and $T$ according to the established intervals. Subsequently, for each randomly chosen pair $(\tilde{J}, \tilde{T})$, we record a single representative spin configuration, generated as follows. After initializing the system in a random spin configuration it is subjected to a process in which it is cooled starting from a predefined high temperature, $T_0 = 5$, down to the sampled target temperature $\tilde{T}$. This particular value of $T_0$ has been chosen to assure that, for any parameter sample $(\tilde{J}, \tilde{T})$, the generation process initializes in the same disordered paramagnetic state common to all Ising-like spin system, minimizing bias in the data set.       

The cooling protocol is implemented using standard single-spin-flip Metropolis Monte Carlo (MC) simulations~\cite{Newman1999} in which the temperature $T$ is reduced at a rate of $2\times 10^{-4}$ per MC sweep, which is defined as a set of $N$ random spin-flip trials such that, on average, each of the $N$ spins in the system is given the opportunity to alter its state. After the cooling stage is completed, the system evolves isothermally at the target temperature $\tilde{T}$ for an additional $3\times 10^3$ MC sweeps, after which the final configuration is recorded in the data set. Since only a single configuration is registered for each randomly sampled parameter pair $(\tilde{J}, \tilde{T})$, all collected spin configurations are statistically independent. 

\subsection{Data-analysis}

Our data-analysis strategy to estimate the critical manifolds is based on three elements. First, we subject the raw MC configurations to unsupervised-learning techniques with the aim of achieving dimensionality reduction of the data. In addition to having shown to be effective in capturing essential features of physical systems~\cite{Wetzel2017,Wang2016}, from a data-analysis standpoint it is useful for tackling difficulties associated with the high-dimensional nature of the raw data set~\cite{Murphy2012}. Next, we process the reduced-space results using clustering algorithms~\cite{Estivill-Castro2002} to identify distinct groups within the data. In some situations dimensionality reduction alone suffices to identify different groups within the data set, but in more complex scenarios, such as in this study, this is not the case. Either way, clustering techniques should be applied to the reduced space so that the identification of coherent groups is unbiased. After this clusterization we map the elements of the identified clusters to the phase diagram using the values of the parameters $\tilde{J}$ and $\tilde{T}$ associated with each data point and verify whether the distinct agglomerates are located in different regions, to be interpreted as distinct phases. Finally, we use the labels generated by the clustering procedure to train a classifier and interpret the obtained decision thresholds in terms of the critical manifolds. Below we describe the details of each of the three elements, all of which have been implemented using the \verb|Scikit-learn| platform~\cite{Pedregosa2011}.

\subsubsection{Dimensionality reduction}

We apply two different approaches for the dimensionality reduction step, using both principal-component analysis (PCA)~\cite{Pearson1901,Jolliffe2006} as well as develop an auto-encoder~\cite{Kramer1991}. PCA achieves dimensionality reduction by determining the eigenvectors, also known as principal components, of the covariance matrix of the raw MC data (with dimensions $N\times N$, with $N$ the number of spins in the system). The principal components (PCs) are then ranked in order of decreasing eigenvalues. The first eigenvector (i.e., that with the largest eigenvalue) corresponds to the high-dimensional direction in the spin space that has the largest variance in the data set. Subsequently, the second eigenvector corresponds to the direction with the second-largest variance, and so on. The assumption then is that only a few PCs are sufficient to capture the essential information contained in the data set and, possibly, also allowing to group them. A restriction of the PCA approach, however, is that it is a fundamentally linear operation, excluding the possibility to detect non-linear relationships among the variables in the data set.

\begin{figure}[t!]
    \centering
    \includegraphics[width=1.\columnwidth]{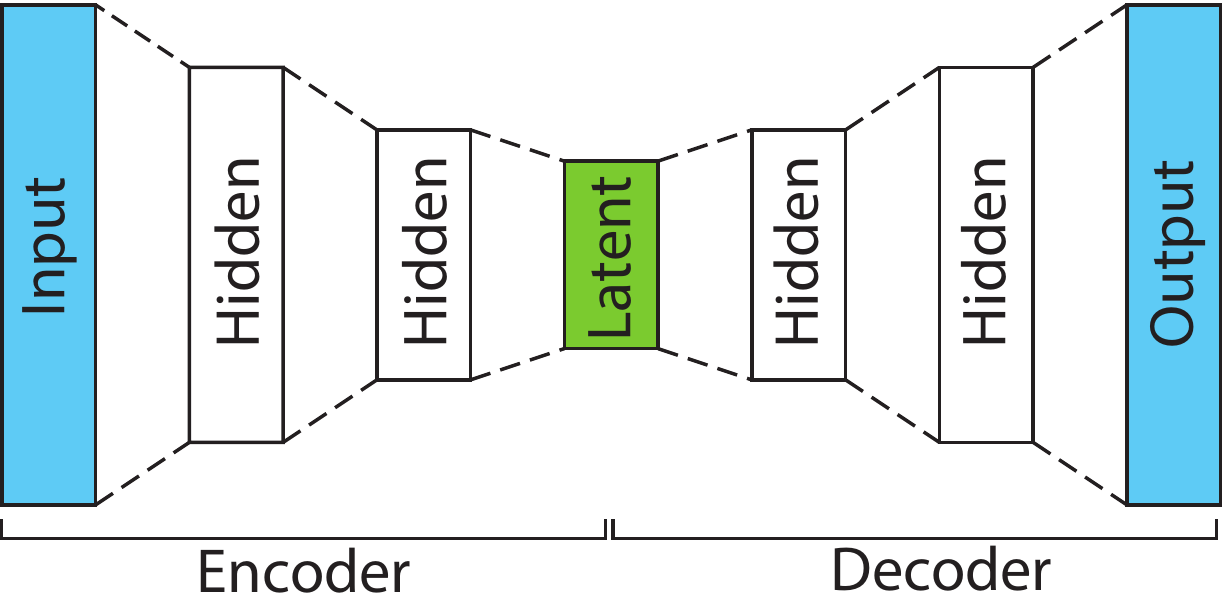}
    \caption{Hourglass structure of auto-encoder neural network. Each block represents a set of neurons. Number of neurons in input and output layers equals the dimension of the data set. Starting from the input layer, the subsequent hidden layers systematically reduce the number of neurons until reaching the bottleneck latent neuron block in which the number of neurons reaches a minimum. Subsequently, the Intermediate, hidden layers, systematically reduce the number of neuron until reaching the latent neuron block in which the number of neurons reaches a minimum. Part between the input and the latent block is referred to as the encoder. Part between the latent block and output is referred to as the decoder.}
    \label{Fig2}
\end{figure}

Auto-encoders, on the other hand, allow detection of such non-linearities. They are neural networks which take the elements of the data set as input and are trained to reproduce that input in the output layer. The structure is that of an hourglass, as shown schematically in Fig.~\ref{Fig2}. Starting from the input layer, with a number of neurons equal to the dimension of the data set, each subsequent hidden layer decreases its number of neurons until reaching the latent neuron block, which is the layer with the fewest neurons. This part of the structure is referred to as the encoder. The other part of the hourglass is known as the decoder, in which the number of neurons in the hidden layers increases again, in a manner symmetric to the encoder part, until reaching the original number of neurons in the output layer. The dimensionality reduction of the data is achieved by the encoder part, with the bottleneck layer spanning the so-called latent space, which represents a nonlinear distilled representation of each input sample. 

Compared to PCA, which amounts to applying a straightforward linear transformation to the data, the creation of an auto-encoder is much more involved. In addition to specifying the number of layers in the hourglass structure and the number of neurons in each of them, including the minimum number of neurons of the latent space, it requires the definition of the neural network. This entails defining the connectivity between the layers, the weights of the connections between the neurons and the functional forms of the activation functions.  

\subsubsection{Clustering}

Among the many available clustering techniques~\cite{Estivill-Castro2002} we use the density-based spatial clustering of applications with noise (DBSCAN) algorithm~\cite{Ester1996}. This choice is motivated by a number of arguments. First, it does not require to pre-define the number of clusters as input. Moreover, it is well-suited to handle noisy datasets and, as opposed to the vast majority of clustering methods, it can identify groups with arbitrary shape. The main principle of the DBSCAN approach is to group together data points that lie within a neighborhood of a specified radius $\varepsilon$. In particular, it searches for those data points that, within this neighborhood, have a specified minimum number of neighbors. Each point that satisfies this criterion is classified as a core point. If a data point does not satisfy this property but it is part of the neighborhood of a core point it is considered a border sample. Finally, if none of these conditions are met, the data point is considered noise. 
\begin{figure}[t!]
    \centering
    \includegraphics[width=0.8\columnwidth]{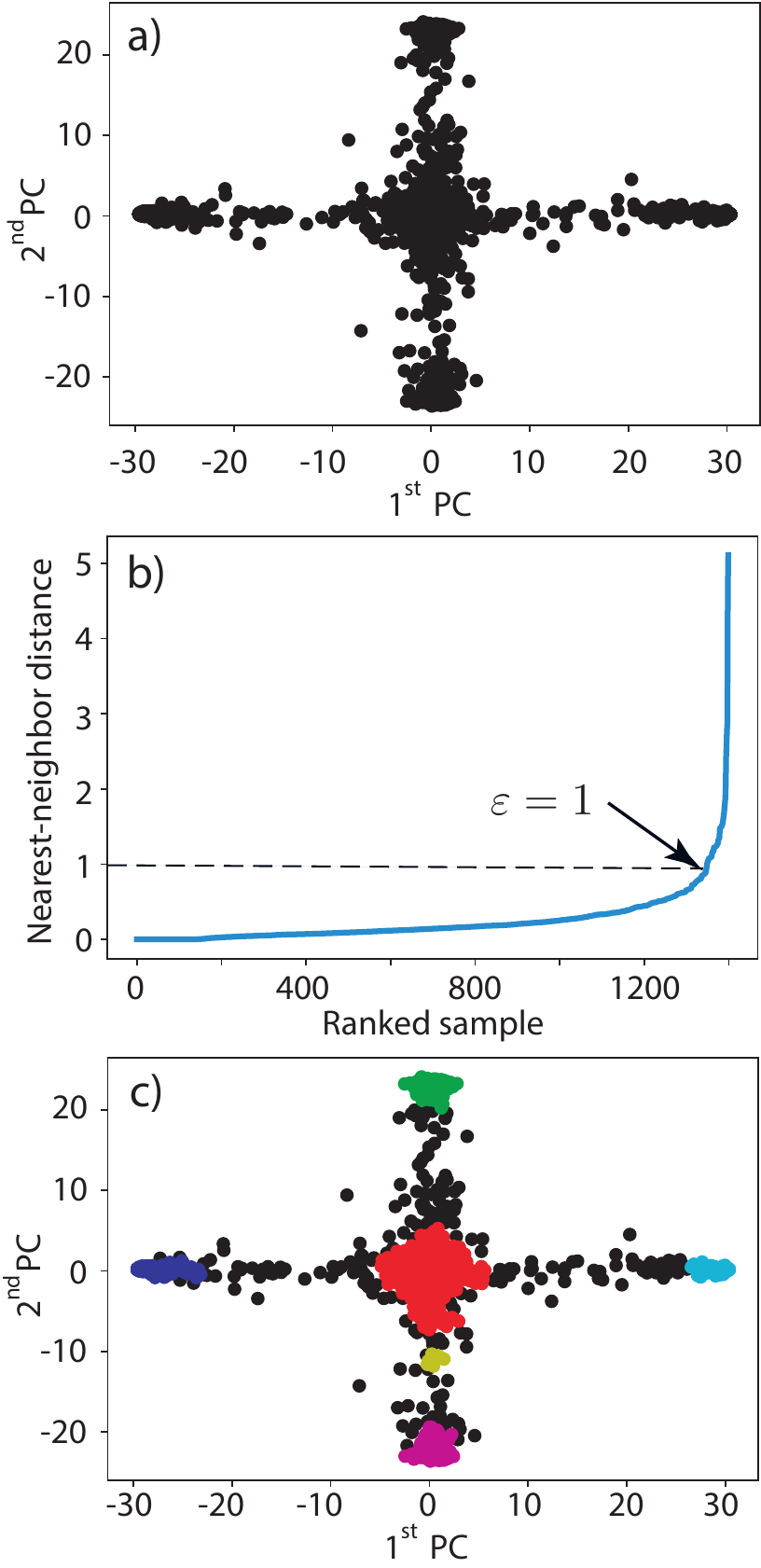}
    \caption{PCA dimensionality reduction of the MC data for the PUD model followed by clustering using DBSCAN. a) Projection of the data on the space spanned by the first 2 PCs. b) Nearest-neighbor distance for each data point in this space, ranked from the smallest to largest. Arrow indicates the location of maximum curvature, which has been shown to provide an appropriate value for the nearest-neighbor radius parameter $\varepsilon$ for the DBSCAN clustering algorithm~\cite{Rahmah2016}. c) Results after clustering. Black circles represent data points classified as noise. Different colors correspond to data points attributed to distinct clusters.}
    \label{Fig3}
\end{figure}

\subsubsection{Classification}

Finally, after the clusterization step, we apply a classification approach to determine the boundaries between different clusters. To this end we use a support-vector machine (SVM)~\cite{Cortes1995}, which provides a robust method for determining nonlinear and fuzzy intersections between clusters.

\section{Results and Discussion}
\label{Results}

The main results for the PUD model are based on a data set containing 1400 independent spin configurations on a 30$\times$30 2D square lattice subject to periodic boundary conditions,
each obtained for a single, uniformly sampled parameter pair $(\tilde{J},\tilde{T})$, as discussed in Sec.~\ref{DataGeneration}. Figure~\ref{Fig3}~a) displays the results obtained after a dimensionality reduction using PCA, with each data point representing one of the 1400 spin configurations as projected on the two-dimensional space spanned by the first two PCs. Next, to apply the DBSCAN clustering approach, we first need to select an appropriate value for the neighborhood radius parameter $\varepsilon$. To this end, we analyze the distribution of nearest-neighbor distances in PCA-reduced space depicted in Fig.~\ref{Fig3}~a). Figure~\ref{Fig3}b) plots the value of this nearest-neighbor distance for each data point, ranked from lowest to highest. It has been shown~\cite{Rahmah2016} that the optimal value of $\varepsilon$ corresponds to the distance at which the curvature of the distance versus rank curve is maximum, which in this case is $\varepsilon \simeq 1$, as indicated by the arrow. Figure~\ref{Fig3}~c) then shows the results after clustering with DBSCAN, using a neighborhood radius $\varepsilon=1$ and setting the minimum number of neighbors within this radius to be 10. Aside from the black circles, which are configurations that have been classified as noise, the data points painted with different colors belong to distinct clusters.   

\begin{figure}[t!]
    \centering
    \includegraphics[width=1.0\columnwidth]{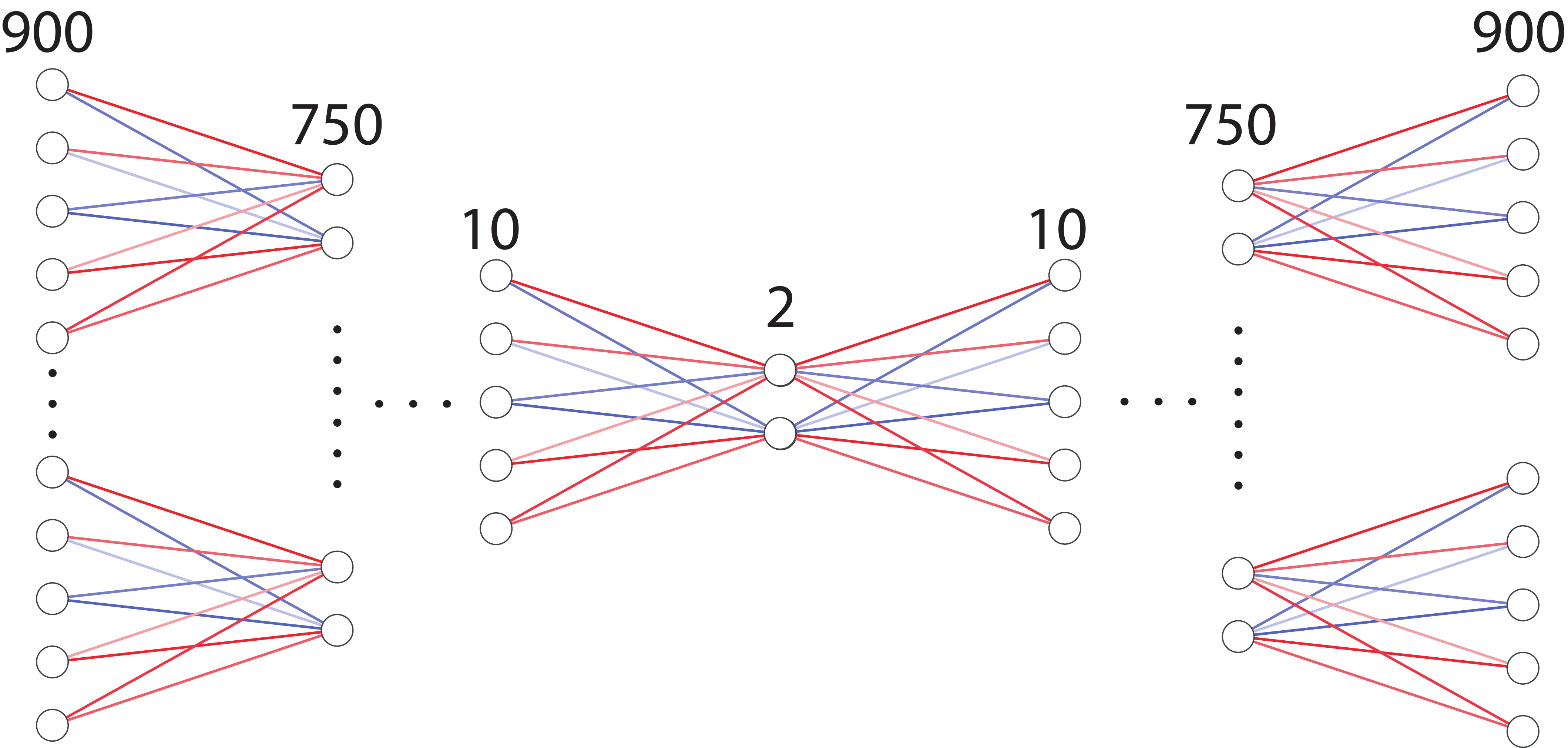}
    \caption{(Color online) Artificial neural network structure of auto-encoder. In addition to the input layer consisting of 900 neurons, the encoder portion consists of 8 hidden layers with, respectively, 750, 600, 450, 300, 150, 75, 30 and 10 neurons, before reaching the latent layer containing 2 neurons. The decoder portion is symmetric with respect to the latent layer. The neural network is linked such that subsequent layers are fully connected.}
    \label{Fig4}
\end{figure}

\begin{figure}[t!]
    \centering
    \includegraphics[width=1.0\columnwidth]{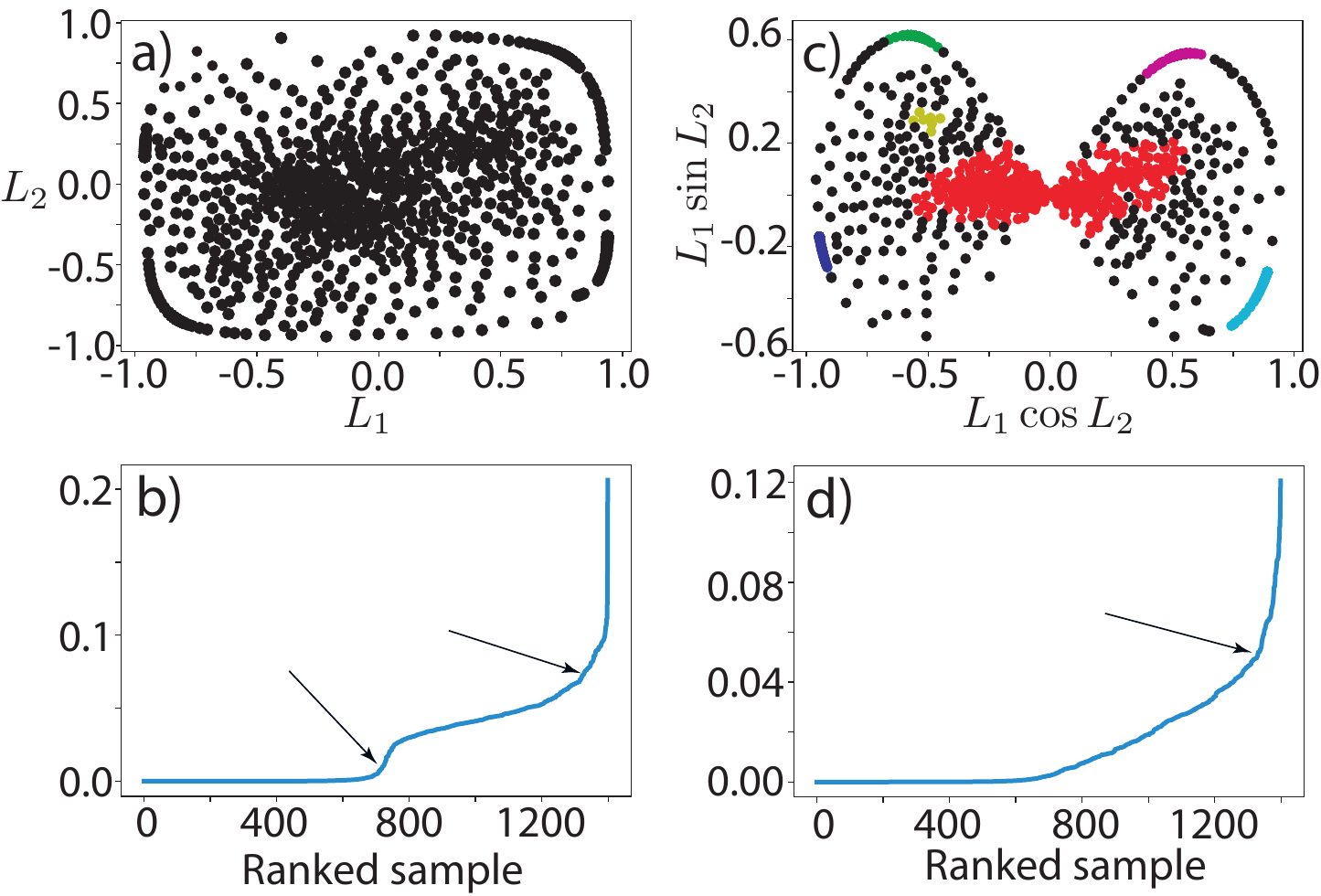}
    \caption{Auto-encoder dimensionality reduction and subsequent DBSCAN clusterization for the PUD data set. a) Data represented in the latent space as characterized by the output $L_1$ and $L_2$ of the corresponding 2 neurons. b) Cluster density analysis based on ranked nearest-neighbor distances for data points shown in a). c) Transformation of the latent space in terms of the transformed variables $L_1 \cos L_2$ and $L_1 \sin L_2$, respectively. Different colors represent data points in distinct clusters as obtained using DBSCAN. Black circles indicate configurations classified as noise. d) Cluster density analysis based on ranked nearest-neighbor distances for latent-space points shown in c).}
    \label{Fig5}
\end{figure}

As a second approach toward dimensionality reduction we develop an auto-encoder with the schematic structure shown in Fig.~\ref{Fig2}. The resulting optimized neural-network structure is depicted in Fig.~\ref{Fig4}. In addition to the 900 neurons on the input layer, the encoder section features a succession of 8 hidden layers with, respectively, 750, 600, 450, 300, 150, 75, 30 and 10 neurons, before reaching the latent layer that consists of 2 neurons. The decoder part is symmetric with respect to the latent layer. The neural network between successive layers is fully connected and we employ hyperbolic tangents as activation functions~\cite{Murphy2012}. The optimization of the auto-encoder neural network was implemented using the \verb|PyTorch| package~\cite{Paszke2019}, employing its MSELoss function~\cite{Goodfellow2016} as the loss measure and the Adam algorithm as the adaptive optimizer~\cite{Goodfellow2016}. The optimization process is organized in three steps. First, the auto-encoder is pre-trained using 400 of the 1400 system configurations at a learning rate~\cite{Goodfellow2016} of $5\times 10^{-4}$ for 1500 iterations (i.e., epochs\cite{Goodfellow2016}). Subsequently, the training procedure covers the entire data set for 4000 more iterations using the same learning rate. Finally, an additional 2000 iterations is carried out for the entire data set at a learning rate of $5\times 10^{-5}$.
The corresponding dimensionality-reduced representation of the data set is shown in Fig.~\ref{Fig5}~a), which depicts the outputs of the neurons $L_1$ and $L_2$ of the latent layer for the 1400 spin configurations. Subsequently, to determine the value of the radius parameter $\varepsilon$ for DBSCAN, we determine the nearest-neighbor statistics of the data set in the same way as done in Fig.~\ref{Fig3}~b). 

The results are shown in Fig.~\ref{Fig5}~b), which shows the nearest-neighbor distance for all data points in Fig.~\ref{Fig5}~a), ranked from lowest to largest. The fundamental difference between this profile and the one obtained for PCA is that, in this case, there are two local maxima in the curvature of the
the rank-distance plot, as shown by the arrows. This indicates that groups belonging to the latent space as represented in Fig.~\ref{Fig5}~a) have different characteristic densities. The existence of two density profiles can in fact be traced to the physical characteristics of the PUD model, as we will discuss later on, but under these conditions the DBSCAN clustering algorithm is known to be less effective. If one chooses  $\varepsilon$ according to the smaller of the two, i.e., focusing predominantly on clustering data points characteristic of high density regions, lower-density samples will not be recognized as belonging to any group at all. In contrast, if we select the larger of the two, i.e., choose cluster according to lower density regions, regions of higher density will incorporate data points that should not be included, giving rise to cluster overlap.

In this case, it has been shown that a transformation of coordinates in the latent space may be helpful to improve the identification of distinct clusters. Specifically, polar-coordinate-like interpretations have shown to particularly useful on a number of occasions, including in the recognition of handwritten digits~\cite{Davidson2018,Joshi2016,Patil2019}. Following this approach, we transform the neuron outputs $L_1$ and $L_2$ of Fig.~\ref{Fig5}a) by interpreting $L_1$ as a generalized ``radius'' (allowing both positive and negative values) and $L_2$ as an ``angle'', giving new Cartesian components defined as $L_1 \cos L_2$ and $L_1 \sin L_2$, respectively. Inverting the roles of $L_1$ and $L_2$ in this transformation does not alter the results. Figure~\ref{Fig5}~c) and d) display the corresponding structure of the transformed latent space and rank-distance curve, respectively, with the latter now having a single point of maximum curvature, allowing an effective application of the DBSCAN clustering algorithm. As in Fig.~\ref{Fig3}~c), the different colored data points in Fig.~\ref{Fig5}c) correspond to distinct clusters identified by DBSCAN, whereas the black circles correspond to configurations that are classified as noise.

\begin{figure}[t!]
    \centering
    \includegraphics[width=0.8\columnwidth]{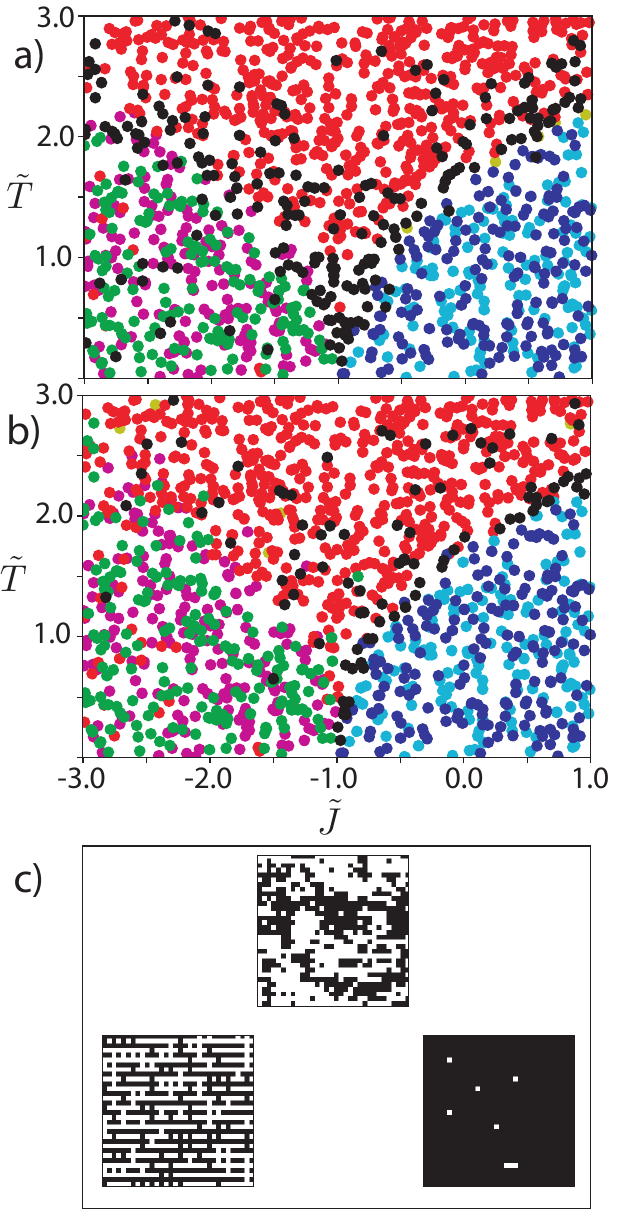}
    \caption{(Color online) Mapping of reduced-space clusters into the PUD parameter space. a) PCA clustered data. b) Auto-encoder clustered data. c) Representative spin configurations for the three phases with black/white squares representing spin up/down, respectively.}
    \label{Fig6}
\end{figure}

Having clustered the data for both dimensionality-reduced representations, we now map the corresponding spin configurations onto the $\tilde{T}\times\tilde{J}$ parameter space of the PUD model, maintaining the color coding adopted in Figs.~\ref{Fig3}~c) and ~\ref{Fig5}~c). The corresponding results are depicted in Figs~\ref{Fig6}~a) and b), which display the mappings produced by PCA and the auto-encoder, respectively. An immediate observation is that, in both cases, each non-noise color occupies only a restricted part of the parameter space, dividing it into three distinct regions. Figure~\ref{Fig6}c) shows representative spin arrangements from these areas, clearly showing the distinct nature of the spin conformations in each of them. In this sense, each region represents a distinct spin phase, displaying a disordered paramagnetic phase, a ferromagnetic phase and a stripe-like anti-ferromagnetic phase. A further notable feature of the data mapping in Fig.~\ref{Fig6}a) and b) is that, while the paramagnetic region corresponds to a single cluster, the ferro and anti-ferromagnetic parts of the parameter space are occupied by two distinct clusters each. We will further discuss this point below.

\begin{figure}[t!]
    \centering
    \includegraphics[width=0.8\columnwidth]{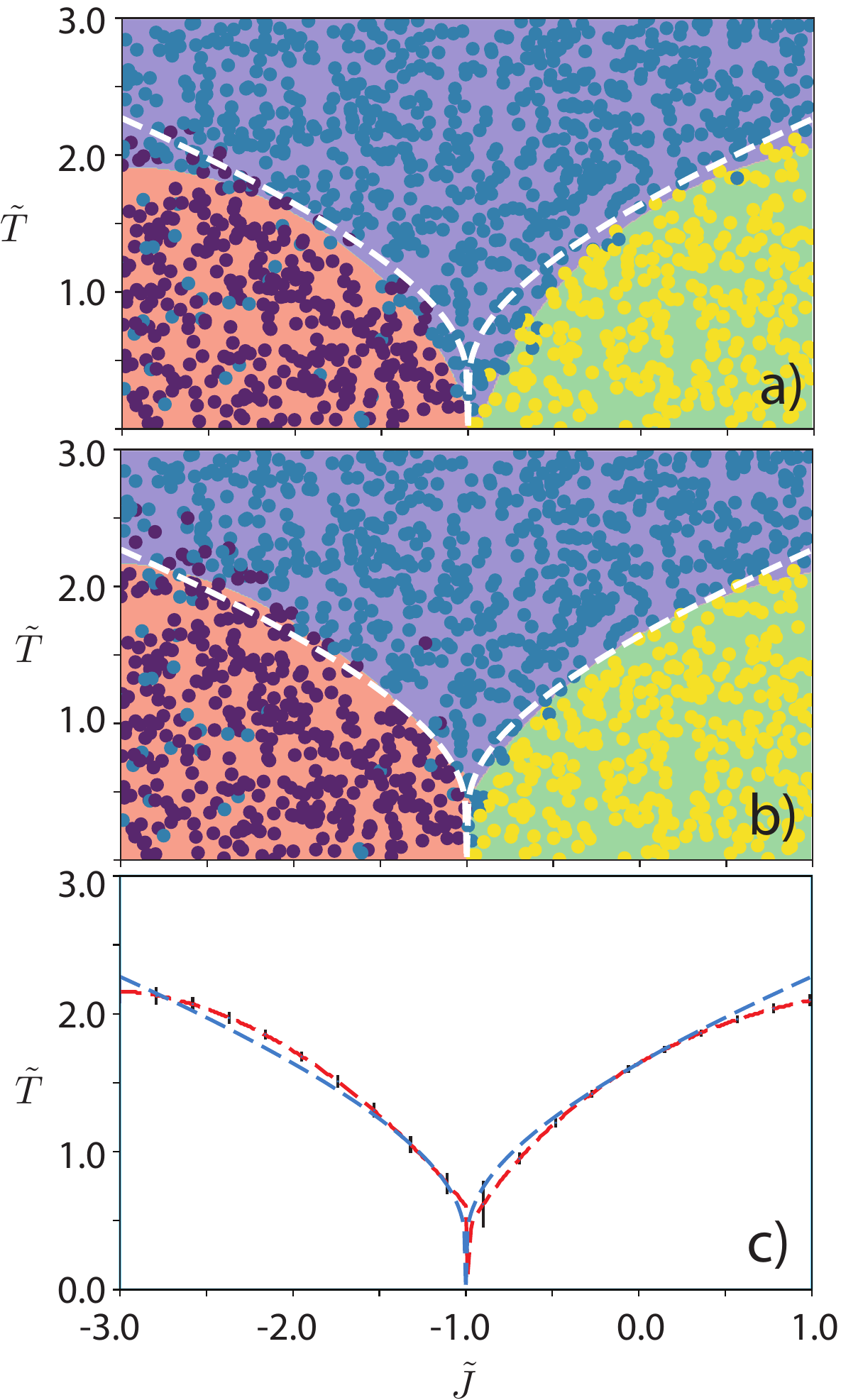}
    \caption{(Color online) Colored circles represent the spin configurations after DBSCAN clustering of the PCA and auto-encoder dimensionality reduction. Dark and light blue data points in Fig.~\ref{Fig6}a) and b) have been assigned to the yellow circles, the green and magenta to purple, and the red and black to blue. Different background colors represent regions defined by the SVM as belonging to a specific class. The boundaries between them represent the decision thresholds. White dashed lines represent analytical critical lines corresponding to the solutions of Eq.~(\ref{eq2}). a) PCA-based phase diagram. b) Polar-coordinate latent-space phase diagram from auto-encoder. c) SVM decision threshold (red line) including uncertainty estimates (vertical black bars) as estimated using the train-test split method~\cite{Pedregosa2011} compared to analytical critical lines (blue line).}
    \label{Fig7}
\end{figure}

Now that the various clusters in the dimensionality-reduced spaces of PCA and the auto-encoder in Fig.~\ref{Fig6} have been associated with different types of spin configurations, we now analyze the data so as to determine the manifolds in the parameter space that separate the different phases. To this end, we define three different classes, corresponding to the three regions in the parameter space identified in Fig.~\ref{Fig6}~a) and b). In particular, the classes consist of, (i), the spin configurations from red cluster for the paramagnetic phase, (ii) the data from the light and dark blue clusters for the ferromagnetic phase and, (iii), the green and magenta clusters for the stripe-like anti-ferromagnetic phase. Based on these classes we employ a supervised-learning SVM multi-classification approach to establish the boundaries between these regions. The corresponding results for PCA and the auto-encoder are shown in Fig.~\ref{Fig7}~a) and b), respectively. The different background colors correspond to the different classes, with the boundaries between them representing the so-called decision thresholds that form the critical manifolds in the parameter space that separate the different phases. We can compare these results directly to the critical lines defined by Eq.~(\ref{eq2}), shown as the dashed white lines.

Overall, both PCA and the auto-encoder closely capture the qualitative features of the critical lines in the parameter space, including the existence of a critical point at zero temperature for $\tilde{J}=-1$. It is clear, however, that the phase-behavior description provided by the auto-encoder is superior in terms of quantitative agreement, with the SVM decision thresholds closely overlapping the analytical results. This is further demonstrated in Fig.~\ref{Fig7}~c), which compares the SVM decision thresholds including confidence intervals as obtained using the train-test split method~\cite{Pedregosa2011}, to the exact results~\cite{Andre1979}. The agreement is encouraging, in particular because the data points were selected randomly and uniformly within a very broad range of the parameter space, without any bias toward known transition lines. 

The main reason for the quantitative difference between PCA and the auto-encoder is the fact that the first 2 PCs cover only $34\%$ of the variance in the data, with the remaining $66\%$ being diluted over the other 897 eigenvectors. The auto-encoder, on the other hand, provides a more comprehensive data reduction scheme. Although it does not involve the concept of explained variance, as for PCA, the fidelity obtained in training the auto-encoder can be used as a measure for the accuracy in the recovery of the input image by the encoder. In this particular case, the auto-encoder can be pushed to reproduce input images with a fidelity superior to 90\%. As discussed previously for a variety of other classical spin systems~\cite{Hu2017,Acevedo2021}, the superiority of the auto-encoder approach is not surprising given that PCA is intrinsically limited to linear transformations of the data.

Even so, the PCA dimensionality reduction, followed by DBSCAN clustering still provides insightful information regarding the physical characteristics of the system. In particular, it has been shown to be useful in the identification of order parameters characterizing phase transitions~\cite{Hu2017,Kiwata2019}. This can be appreciated in the cluster structure depicted in Fig.~\ref{Fig3}~c), in which vertically and horizontally opposite clusters belong to the same striped anti-ferromagnetic and ferromagnetic phases, respectively, whereas the centralized red agglomerate corresponds to the paramagnetic phase. In this sense, the first PC clearly distinguishes between the two symmetric ferromagnetic states (i.e., spin up and spin down) and the paramagnetic phase, whereas the second PC does so to differentiate the (spin-up and spin-down) striped anti-ferromagnetic phases from the paramagnetic phase.

\begin{figure}[t!]
    \centering
    \includegraphics[width=0.8\columnwidth]{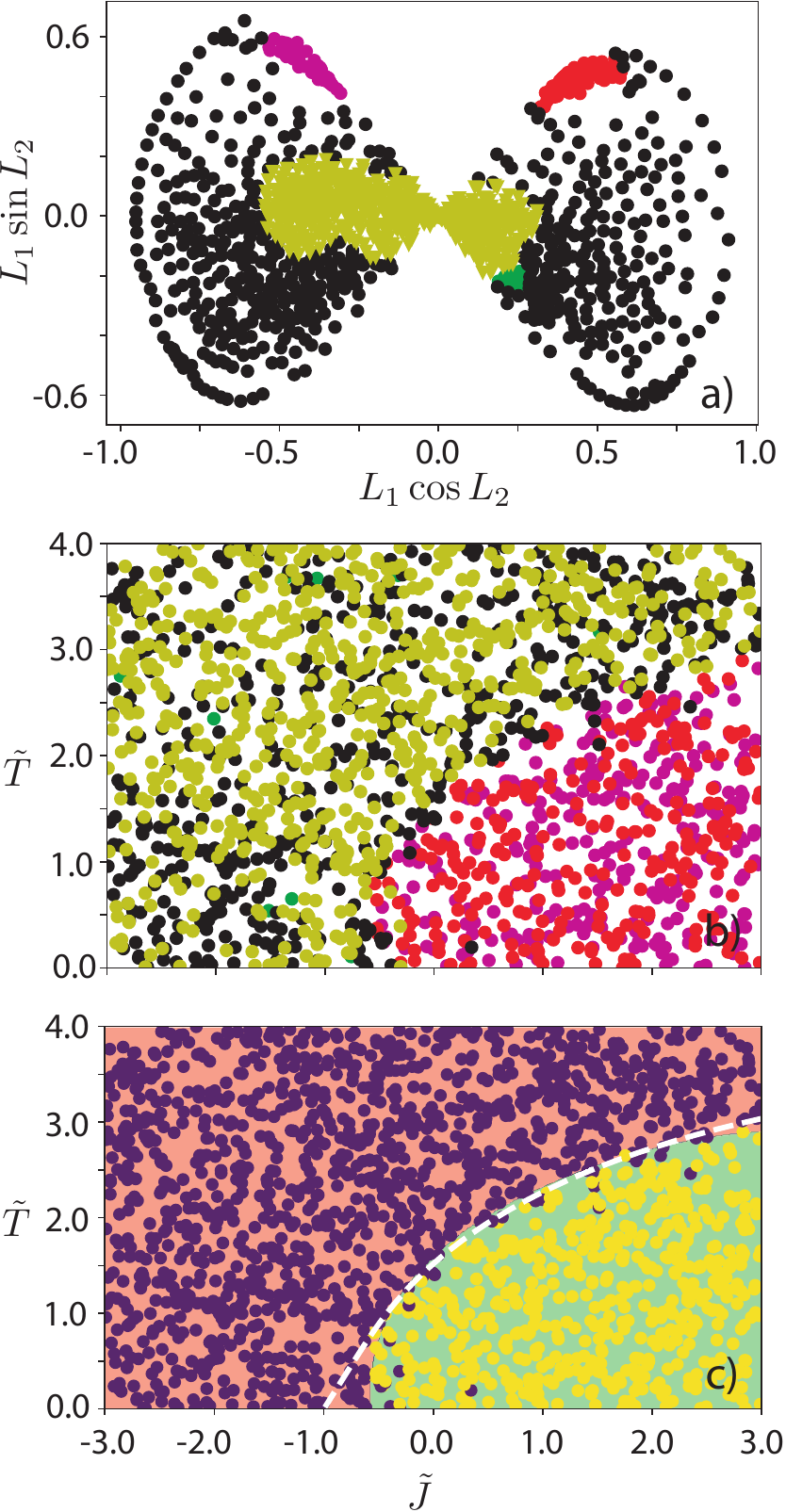}
    \caption{(Color online) Auto-encoder results for ZZD model. a) Hyperspherical latent space after  DBSCAN clustering, with different clusters represented by different colored circles. b) Spin configurations plotted in the $(\tilde{J},\tilde{T})$ parameter space, maintaining the colors from a). c) Corresponding ZZD phase diagram as obtained by SVM classification. Different background colors represent regions defined by the SVM as belonging to a specific class. The boundaries between them represent the decision thresholds. White dashed line represents analytical critical line from Eq.~(\ref{eq3}). }
    \label{Fig8}
\end{figure}

Fig.~\ref{Fig8} shows the results obtained for the other frustrated spin system, defined by the ZZD model, which is characterized by a different geometric patterns of the coupling parameters $J$ and $J^{\prime}$. The dimensionality reduction is obtained using an auto-encoder with the same two-neuron bottleneck architecture applied for the PUD model. Fig.~\ref{Fig8}~a) displays the resulting latent-space structure after applying the same polar-coordinate-like transformation used for the PUD model and using DBSCAN clusterization. It displays the same butterfly-like shape as for the PUD, but in this case the number of distinct clusters is reduced by two. Specifically, the two mirror-symmetric clusters on the lower part of the butterfly for the PUD model have disappeared. When mapping the data points onto the $(\tilde{J},\tilde{T})$ parameter space, as depicted in Fig.~\ref{Fig8}~b), it is clear this reduction is related to the fact that the ZZD model is characterized by two instead of three phases, displaying a ferromagnetic phase for positive values of $\tilde{J}$ and low $\tilde{T}$ and a disordered paramagnetic phase for negative values of $\tilde{J}$ or high temperatures. After grouping the clusters belonging to the same region of the phase diagram into distinct classes and training an SVM classifier we obtain the critical line separating both phases as the boundary separating both background colors, as shown in Fig.~\ref{Fig8}~c). As for the PUD model, the agreement with the exact result described by Eq.~(\ref{eq3}), shown as the white dashed line, is excellent. 

An interesting question is to what extent there is a relation between the structural characteristics of the optimized auto-encoder and the fundamental physical characteristics of the two spin models. In this context, we note that for both spin models the best results are obtained using a bottleneck layer formed by 2 neurons, obtaining a reproduction fidelity $\gtrsim 90$\%. If, for instance, only a single neuron is used, the optimized fidelity does not exceed $\sim 50$\%, whereas a bottleneck layer containing three neurons does not significantly improve the fidelity. The fact that the optimal latent-space dimension is two for both systems, even though the number of distinct phases for both models is different (3 for the PUD versus 2 for the ZZD), suggests that the optimal dimension of the latent space may be related to the number of independent parameters required to characterize the phase diagram. A further indication for the connection between the structure of the auto-encoder latent space and the physical characteristics of the system can be seen in Figs.~\ref{Fig5}~c) and \ref{Fig8}~a). The butterfly-shaped hyper-spherical latent spaces are mirror-symmetric with respect to the vertical axis passing through the origin, with the symmetry-related cluster pairs belonging to the same regions in the phase diagram. The symmetry brought out by the clustered auto-encoder results thus provides insight into the physical system under consideration, revealing that opposite clusters contain configurations that are spin reversals of each other. In other words, the vertical axes in Figs.~\ref{Fig5}~c) and \ref{Fig8}~a) correspond to the spin-reversal symmetry underlying the system Hamiltonians. This ``discovery'' of a fundamental system symmetry through the optimization of the auto-encoder is a further example that the global ML phase exploration employed here can be useful in perceiving fundamental properties of physical systems when prior theoretical insight is unavailable.   

In a similar context, the existence of two density profiles in the Cartesian auto-encoder latent space for the PUD model, as shown in Fig.~\ref{Fig5}~b), is in fact related to the nature of its ferro and antiferromagnetic phases of the PUD model. Theoretical analysis~\cite{Andre1979} indicates that the entropy of its anti-ferromagnetic phase is much greater, meaning that the variability among different samples is much larger compared to that for the ferro-magnetic configurations. This is the reason underlying the existence of the two density profiles for the ranked sample-sample distances in Fig.~\ref{Fig5}~b). Whereas the clusters belonging to the lower $\varepsilon$ (i.e., higher density) correspond to ferromagnetic configurations, the ones for the larger $\varepsilon$ value represent anti-ferromagnetic spin conformations. Even though the data does not allow quantification of the configurational entropy for both phases, it does provide qualitative indications that it is larger for the striped anti-ferromagnetic phase, without previous knowledge of the system. 

It is interesting to contrast the present fully-connected auto-encoder approach to that employed for the two-dimensional frustrated antiferromagnetic Ising model as reported in Ref.~\citenum{Acevedo2021}. In the latter, a convolutional auto-encoder (CAE) was used to determine  the critical line between the antiferromagnetic and paramagnetic phases. The advantage of CAEs over the fully connected auto encoder utilized in the present work is that they are characterized by a much smaller number of parameters and feature the ability to recognize spatial correlations by the application of filters or local transformations on the images, allowing, for instance, to construct the phase diagram of the Bose-Hubbard model~\cite{Kottmann2020}. Even so, in addition to the fact that the implementation of CAEs is generally substantially more involved, the CAEs in Ref.~\citenum{Acevedo2021} do not have a low-dimensional latent space, such that its detection of phase transitions is not based on dimensionality reduction, but rather on anomaly detection. In this sense, both auto-encoder techniques approach the construction of phase diagrams in very different manners and may be considered as complementary to each other. Particularly useful features of the present fully-connected auto-encoder approach are its interpretability in terms of the latent space features and its efficiency, giving accurate estimates for the critical lines based on only 1400 MC spin configurations.

A final point concerns the influence of the system size on the global estimate of phase diagrams. In applications of ML techniques used to locate a single critical point, such a size-dependence analysis has been instrumental in achieving quantitative accuracy~\cite{Carrasquilla2017,Hu2017}. In the context of classical spin systems, for instance, such an analysis involves generating configurations in the vicinity of a critical point for different linear lattice dimensions $L$, after which pertinent outcomes are analyzed as a function of some power of $L$. For instance, using PCA for the standard 2D ferromagnetic Ising model, its critical temperature can be determined by plotting the temperature for which the quantified second principal component reaches its maximum as a function of $L^{-1}$, followed by extrapolation to the limit of infinite lattice dimension~\cite{Hu2017}.

\begin{figure}[t!]
    \centering
    \includegraphics[width=1.0\columnwidth]{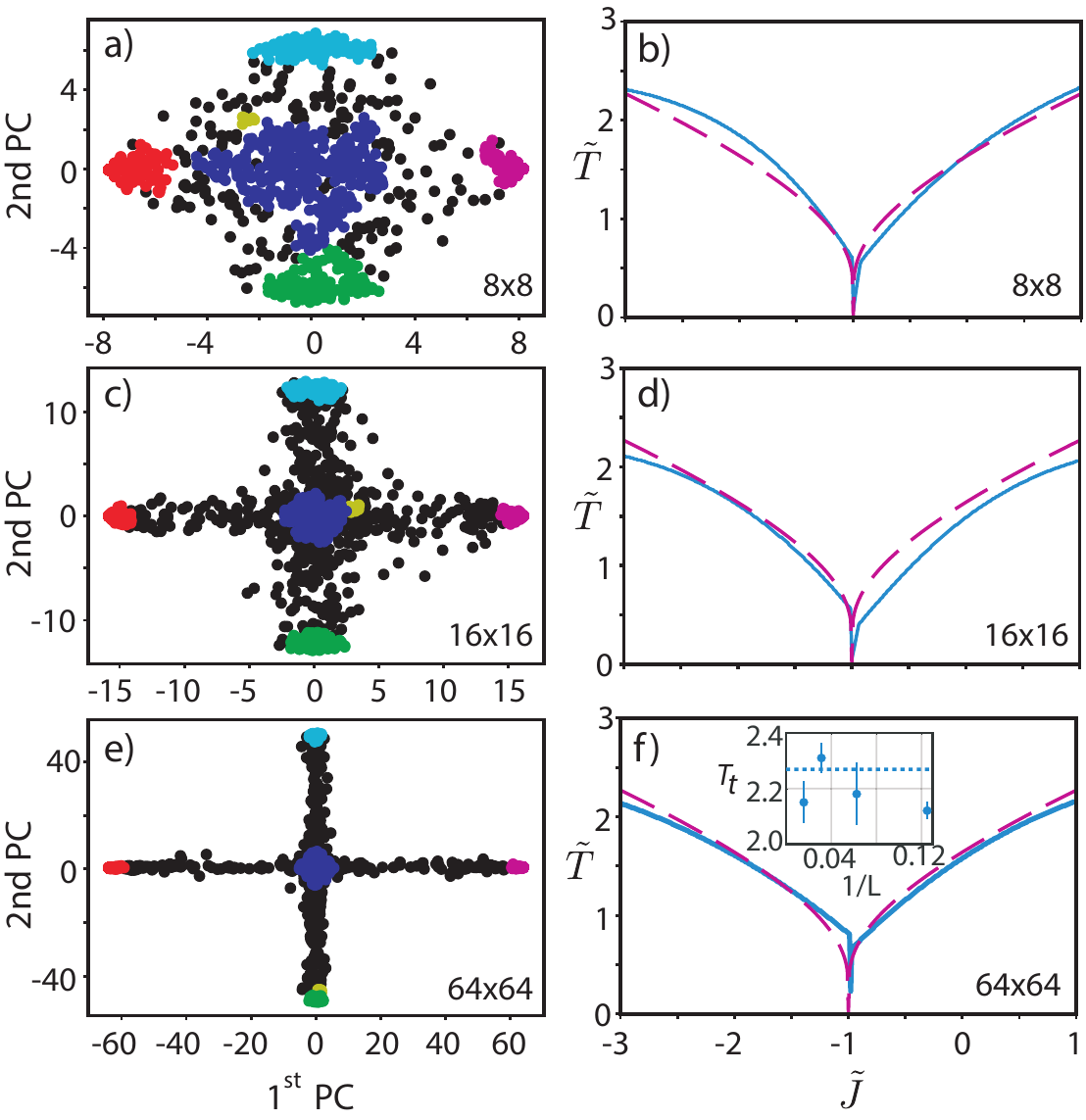}
    \caption{(Color online) System size dependence of latent space and corresponding SVM estimates for the critical lines for PUD model for 8x8 16x16 and 64x64 lattice sizes. Inset in f) displays SVM critical temperature estimate as a function of $L^{-1}$ for $\tilde{J}=1$, with horizontal dashed line indicating exact value $T_c=2/\ln(1+\sqrt{2})$. }
    \label{Fig9}
\end{figure}

In the present global ML approach, however, in which one uses a single data set consisting of configurations sampled from the entire parameter space of the system, local scaling properties become blurred. First, the dimensionality reduction encodes the characteristics of the phase diagram as a whole, incorporating a continuum of critical temperatures instead of a particular single value. Furthermore, in a similar fashion, the final estimates for the critical lines are obtained using a SVM classifier, whose result corresponds to the minimum of a classification error that is global in nature rather than reflecting behavior in the vicinity of a single critical point. 
This is illustrated in Fig.~\ref{Fig9}, which displays the PCA results for the PUD model for lattice dimensions $L=8$, \, 16 and 64. The size effect on the distribution of the data points in the clustered latent space is evident. Whereas the results display appreciable dispersion for $L=8$, the projections onto the first two principal components become progressively sharper as the system size grows, increasing the distinguishability among the ferromagnetic, anti-ferromagnetic and paramagnetic phases. On the other hand, even though the SVM critical lines for $L=64$ are visibly better than those obtained for $L=8$ in an overall sense, there is no manifestly visible size-scaling behavior. This is shown in the inset of Fig.~\ref{Fig9}~f), which plots the SVM critical temperature as a function of $1/L$ for the case of the standard 2D Ising model (i.e., $\tilde{J}=1$). Even though the estimates are within 7\% of the exact value for all system sizes, there is no discernible $L^{-1}$ dependence. 

In this sense, the present global ML approach can be considered to be complementary to the schemes that have been utilized to quantify the parameters characterizing a particular critical point~\cite{Carrasquilla2017,Hu2017}. Whereas the present approach allows one to obtain a good first picture of the overall phase diagram when no \emph{a priori} insight is available, the latter can subsequently be used to systematically obtain better accuracy for the critical parameters by targeting the generation of additional data in the vicinity of the initial transition estimate and using techniques as described in Refs.~\citenum{Carrasquilla2017} and \citenum{Hu2017}. This is also consistent with the findings of Th{\'e}veniaut and Alet for the many-body localization transition in the Heisenberg spin 1/2 chain~\cite{Theveniaut2019}, who reported that uncertainties in the estimates for critical parameters extracted from neural-network structures tend to be larger than those obtained from conventional approaches.

\section{Conclusions}
\label{Conclusions}

In summary, we consider a ML approach for obtaining a global prediction of the phase diagrams for two frustrated Ising models, which are characterized by the presence of critical lines rather than a single critical point. Using raw MC spin configurations generated for random parameter values from the entire parameter space, we first apply unsupervised learning techniques such as PCA and auto-encoders to achieve dimensionality reduction, followed by clustering using the DBSCAN method and a classification step using a SVM approach. The resulting estimates for the critical lines of both systems are in excellent agreement with available exact results, even though the data points were selected randomly and uniformly within an ample region of the parameter space, without any predisposition toward the critical lines. 

In both cases, the auto-encoder dimensionality-reduction approach is found to provide better quantitative results as compared to those based on PCA. The main origin of this difference is that the optimized auto-encoder gives an image fidelity $\gtrsim 90$\%, with two neurons in the bottleneck layer. In contrast, the corresponding two-dimensional latent space for PCA captures only ~34\% of variance in the data. A further observation is that the latent-space characteristics of the optimized auto-encoders appear to relate to fundamental physical characteristics of both considered spin models. In addition to recognizing the spin-up/spin-down symmetry, in both cases the best results are obtained using a bottleneck layer formed by 2 neurons, indicating that the optimal latent-space dimension is linked to the number of independent parameters required to characterize the phase diagram. Although this relation between the optimal auto-encoder structure and the physical properties requires further study, it suggests that the present ML approach may be helpful in perceiving fundamental physical properties when, for instance, \emph{a priori} theoretical insight is unavailable.   

Finally, although the approach provides good estimates for the phase diagram as a whole, it is not naturally suited to be used in finite-size scaling procedures that are often employed to accurately quantify critical parameters. This is due to the fact that the predictions are based on data sets generated for the entire parameter space rather than in the vicinity of a particular critical point. In addition, the estimates for the transition lines are obtained using a classification procedure that seeks to minimize a global classification error across the phase diagram as a whole such that local critical properties become obscured by global optimization criteria. In this sense, the present approach is complementary schemes that are utilized to quantify critical parameters by focusing on data generated in the vicinity of a critical point.

\section*{CRediT authorship contribution statement}

D.R.A.E: Conceptualization, execution of computations, writing. E.G: Conceptualization, review and editing. M.K.: Conceptualization, writing and supervision. 

\section*{Declaration of competing interest}
The authors declare that they have no known competing financial interests or personal relationships that could have
appeared to influence the work reported in this paper.

\section*{Acknowledgments}
The authors acknowledge support from CNPq, Fapesp grants no. 2016/23891-6 (MK) and 2018/19586-9 (EG), as well as the Center for Computing in Engineering \& Sciences - Fapesp/Cepid no. 2013/08293-7 (MK). We gratefully acknowledge Rodolfo Paula Leite, Levy Boccato, Romis Ribeiro de Faissol Attux and Gabriel Reis Garcia.

\bibliographystyle{elsarticle-num-names} 

\end{document}